\documentclass[floatfix,%
 reprint,
 amsmath,amssymb,
 aps, pra
]{revtex4-1}

\usepackage{physics}
\usepackage{amsmath, amsthm, amssymb,commath, calrsfs, wasysym, dsfont ,verbatim, bbm, color, graphicx}
\usepackage{xcolor}
\usepackage{graphicx}
\usepackage{dcolumn}
\usepackage{bm}
\usepackage[margin=0.95in]{geometry}

\begin{document}

\preprint{APS/123-QED}

\title{Tailored multipolar optical trapping with higher-order structured light}

\author{Nabendu Mishra}
\affiliation{The Institute of Optics, University of Rochester, Rochester, New York 14627, USA}
\author{Nick Vamivakas}
\affiliation{The Institute of Optics, University of Rochester, Rochester, New York 14627, USA}
\affiliation{Department of Physics, University of Rochester, Rochester, NY 14627, USA}
\affiliation{Center for Coherence and Quantum Optics, University of Rochester, Rochester, NY 14627, USA}
\author{Eileen Otte}
\email{eotte@ur.rochester.edu}
\affiliation{The Institute of Optics, University of Rochester, Rochester, New York 14627, USA}
\vspace{3em}
\date{\today}

\begin{abstract}
Dipolar optical forces can confine Rayleigh-sized dielectric particles towards or away from intensity maxima, commonly known as bright and dark trapping. Structured light illumination opens up opportunities for customized trapping landscapes, unveiling the possibility of gray trapping, especially for larger Mie particles that can excite a higher-order multipolar response. While previous studies have primarily focused on calculations of the gradient force for some exemplary, focused scalar beams, we not only extend these analyses by combining vectorial focusing with generalized Lorenz–Mie theory (GLMT), but also consider more complex trapping fields. Motivated by their unique tight focusing behavior, shaping strong transverse intensity gradients and three-dimensional electric field components, we explore higher-order cylindrical scalar and vector beams. We theoretically investigate multipolar trapping of Mie-regime silicon nanospheres in water, calculating the three-dimensional net radiation force and assess local trapping stability through the force Jacobian. Near an octupolar Mie resonance, we predict stable gray trapping with enhanced efficiency for vector beams. Further, we find field geometries enabling configurable, particle-size-dependent gray/dark trapping with subwavelength inter-particle spacing. These results establish polarization, illumination structure and particle parameters as complementary controls for radius-dependent multipolar trapping, providing a basis for potential applications in sub-wavelength particle sorting, trapping-assisted nanopatterning and optical-binding investigations.


\end{abstract}

\maketitle

\section{Introduction}\label{sec1}
Optical trapping uses momentum transfer from light to confine and manipulate micro- and nanoscopic particles, as established by Arthur Ashkin in 1986 \cite{Ashkin:86}. Owing to their ability to exert and measure forces without mechanical contact, optical traps have become versatile, Nobel prize-awarded tools in a vast range of disciplines including cellular biophysics, soft-matter physics, stochastic thermodynamics, neutral-atom quantum science, and levitated optomechanics\,\cite{gieseler2021optical,volpe2023roadmap,millen2020optomechanics}. In the past decades, the marriage between optical trapping and structured light has added a new dimension to optical control and particle manipulation and
led to a deeper understanding of light-matter interactions\,\cite{otte2020optical,yang2021optical,kritzinger2022optical}. In the electric-dipole limit, a particle with positive real electric polarizability experiences a gradient force directed towards increasing intensity, providing the familiar basis for bright trapping\,\cite{chaumet2000time}. Structured light\,\cite{forbes2021structured,otte2020structured} extends control over this interaction through spatial modulation of amplitude, phase, and polarization. For instance, scalar structured beams with uniform polarization and cylindrical vector beams (CVBs) with spatially varying polarization offer distinct ways to construct tailored focal fields\,\cite{Otte2017FocalLandscapes,youngworth2000focusing,otte2020optical}, thereby facilitating the precise customization of optical forces and trapping landscapes tailored to specific application requirements.

In addition to the spatial structure of the electromagnetic field, a crucial degree of freedom for engineering optical control comes from the particle parameters. For instance, high refractive index dielectrics like silicon, which support multiple electric and magnetic Mie resonances, have demonstrated the ability to enhance confinement and enable dark trapping at optical intensity minima\,\cite{lepeshov2023levitated,lu20263d}. In a related study\,\cite{zhang2024full}, Zhang et al. introduced full-gray optical trapping of silicon spheres and distinguished it from bright (intensity maxima) and dark (intensity minima) trapping. Such gray or dark trapping could reduce thermophoretic and heating effects on the particle since the particle accesses regions of moderate or minimum intensity\,\cite{braun2013optically,liu2021opto}. They attributed this behavior to higher-order multipolar resonances that couple nonlocal intensity inhomogeneity to the optical gradient force. Their results utilize a scalar structured illumination, exemplifying the potential capabilities of such fields for trapping in the transverse plane while the prospects of axial trapping still have to be explored. For certain applications, such as in levitated optomechanics, trapping in the axial direction is crucial and needs substantial investigation. Furthermore, the role of polarization in structured trapping landscapes remains unexplored in the context of multipolar optical gray and dark trapping. Tightly focusing higher-order CVBs of spatially varying polarization redistributes the electric field among transverse and longitudinal focal components, which could lead to sophisticated intensity landscapes with exotic force distributions\,\cite{Otte2017FocalLandscapes,GomezViloria2024VortexTrapping,youngworth2000focusing}. 

In this work, we investigate the capabilities of cylindrical scalar and vector beams for multipolar trapping beyond conventional bright trapping. Establishing such trapping landscapes requires evaluating the complete radiation-force balance, rather than only studying the gradient force distribution. For this purpose, we utilize Generalized Lorenz-Mie theory (GLMT), which provides such a framework for calculating the net force on a spherical particle illuminated by an arbitrary beam\,\cite{barton1989theoretical,nieminen2010approximate,kiselev2016optical}. Moreover, optical forces in structured fields can be non-conservative, making the force field and its local path integral an essential tool in our study for assessing the stability of equilibrium positions\,\cite{kiselev2016optical}. 
In particular, we theoretically investigate the optical-force landscapes of silicon spheres illuminated by tightly focused higher-order structured beams in water. We examine particle radii near an octupolar resonance and distinguish gray and dark trappings for various input structured beams. Stable trapping sites are predicted and quantified via the local force Jacobian and its local path integral\,\cite{kiselev2016optical}. We introduce flower- and web-polarization structures as input CVBs and demonstrate their enhanced capability for engineering optical trapping landscapes compared with the previously reported scalar cylindrical beams\,\cite{zhang2024full}. Through axial force calculations GLMT facilitates a complete three-dimensional analysis of the trapping landscape and to investigate the potential for axial trapping. Our approach allows us to engineer particle trapping landscapes that could find practical applications such as particle sorting\,\cite{bobkova2021optical,kong2025customized} with nanometer-level accuracy, optical trapping assisted nanopatterning in the mesoscopic regime \cite{mcleod2008subwavelength,marago2013optical,trompoukis2012photonic}, or drive fundamental research, e.g., studying optical binding forces\,\cite{shukla2026multipolar,taylor2009optical,yan2013guiding,forbes2020optical} under enhanced scattering conditions.

\section{Theory}\label{sec2}
The dipolar model of light-matter interaction fails to correctly capture the gradient force experienced by a Mie scatterer, i.e., when the size of the scatterer is comparable to the wavelength of light $\lambda$. A more accurate description of the gradient force acting on a Mie particle in a focused EM field includes the contributions of the higher-order multipoles, e.g., electric and magnetic quadrupoles and octupoles, and so on, and is given by\,\cite{zhang2024full}
\begin{equation}
    \textbf{F}_G^{(N)}=\sum_{l=1}^NC_{N,l}\left( k^2+\frac{\nabla^2}{2} \right)^{l-1}\nabla|\textbf{E}|^2,
    \label{eq: gradient-force}
\end{equation}
where $N$ denotes the highest order of multipole considered, henceforth referred to as the truncation order, $k$ the wavenumber in the medium, and $\textbf{E}$ the electric field at the particle's center-of-mass (COM) coordinates. The coefficients $C_{N,l}$, are derived from the Mie $a$ and $b$ coefficients, and depend on the multipolar order $l$ and the particle’s refractive index $n$\,\cite{Rahimzadegan:20}. The presence of $\nabla^2$ for $N>1$ alludes to the critical role played by the higher-order multipoles in generating a nonlocal intensity gradient in the vicinity of the particle's COM, resulting in the so-called gray trapping. The gradient force alone, however, does not dictate the force landscape in the plane of optical trapping. As the particle size approaches Mie regime, scattering forces arising from radiation pressure become stronger and need to be accounted for in the picture\,\cite{Ashkin:86,harada1996radiation}. 

We start with a heuristic approach, similar to the works by Harada et al.\,\cite{harada1996radiation}, Bradshaw et al.\,\cite{bradshaw2017manipulating}, and Chaumet et al.\,\cite{chaumet2000time}, where the total radiation force exerted on the particle by the EM field is given by the sum of the gradient and scattering forces. The gradient force considered here is the aforementioned multipolar $\textbf{F}_G^{(N)}$ given by Eq.~(\ref{eq: gradient-force}). The scattering force $\textbf{F}_S$ derived from the radiation pressure cross-section $C_\textrm{pr}$ for an isotropic particle\,\cite{bohren2008absorption}, and the time-averaged Poynting vector $\langle \textbf{S}(\textbf{r},t)\rangle$ corresponding to the particle's COM $(\textbf{r},t)$, can be expressed as 
\begin{equation}
    \textbf{F}_S=\frac{n_\textrm{med}C_\textrm{pr}\langle \textbf{S}(\textbf{r},t)\rangle}{c}.
    \label{eq: scattering-force}
\end{equation}
This results in a simple binomial representation of the total force 
\begin{equation}
    \textbf{F}_\textrm{tot}=\textbf{F}_G^{(N)}+\textbf{F}_S.
    \label{eq: grad_plus_scatt_force}
\end{equation}
While the binomial force model is useful in the Rayleigh scattering regime (particle radius, $R<\lambda/20$ where the dipole approximation is sufficient)\,\cite{kerker2013scattering}, it still fails to agree with the experimental results for a particle in the Mie scattering regime ($R\sim\lambda/(2\pi)$) under non-uniform illumination\,\cite{harada1996radiation,nieminen2010approximate}. This calls for further generalization of the theory, such that it captures the physics of multipolar excitations in a particle of larger scattering cross-section than a Rayleigh-sized particle, and can accurately describe the optical forces not only at the COM but across the surface of the particle\,\cite{yan2007radiation,lock2004calculation} experiencing a non-uniform (structured) intensity landscape. The most widely used approach to solve this challenge is the GLMT for arbitrary focused laser beams\,\cite{barton1989theoretical}. Consider an electric field $\textbf{E}^{(i)}$ incident on a particle of radius $R$ and refractive index $n$ in the focal plane, obtained by tight focusing\,\cite{novotny2006principles} of an arbitrary input beam $\textbf{E}_\textrm{in}$. Following the theory outlined by Barton et al.\,\cite{barton1989theoretical}, the coefficients $A_{lm}$ and $B_{lm}$ in spherical coordinates $(r,\theta,\phi)$, describing the incident electric and magnetic fields on a particle of radius $R$, are given by 

\begin{align}
A_{l m}=&\frac{1}{l(l+1)\psi_l(\alpha)}\notag\\
&\times\int_0^{2\pi}\!\int_0^\pi
\sin\theta\,E_r^{(i)}(R,\theta,\phi)\,Y_{l m}^{*}(\theta,\phi)\,d\theta\,d\phi,
\end{align}
\begin{align}
B_{l m}=&\frac{1}{l(l+1)\psi_l(\alpha)}\notag\\
&\times\int_0^{2\pi}\!\int_0^\pi
\sin\theta\,H_r^{(i)}(R,\theta,\phi)\,Y_{l m}^{*}(\theta,\phi)\,d\theta\,d\phi,
\end{align}
where $\psi_l$ are the Riccati-Bessel functions of the first kind, $Y_{lm}(\theta,\phi)$ are the spherical harmonics, and $\alpha=kR$ is the particle size parameter. The scattered field coefficients $a_{lm}$ and $b_{lm}$ are obtained from $A_{lm}$ and $B_{lm}$ using the equations
\begin{align}
a_{l m}=
&\frac{\psi_l'(\bar n\alpha)\psi_l(\alpha)-\bar n\psi_l(\bar n\alpha)\psi_l'(\alpha)}
{\bar n\psi_l(\bar n\alpha)\xi_l^{(1)\prime}(\alpha)-\psi_l'(\bar n\alpha)\xi_l^{(1)}(\alpha)}
A_{l m},\\
b_{l m}=
&\frac{\bar n\psi_l'(\bar n\alpha)\psi_l(\alpha)-\psi_l(\bar n\alpha)\psi_l'(\alpha)}
{\psi_l(\bar n\alpha)\xi_l^{(1)\prime}(\alpha)-\bar n\psi_l'(\bar n\alpha)\xi_l^{(1)}(\alpha)}
B_{l m},\\
c_{l m}=
&\frac{\xi_l^{(1)\prime}(\alpha)\psi_l(\alpha)-\xi_l^{(1)}(\alpha)\psi_l'(\alpha)}
{\bar n^2\psi_l(\bar n\alpha)\xi_l^{(1)\prime}(\alpha)-\bar n\psi_l'(\bar n\alpha)\xi_l^{(1)}(\alpha)}
A_{l m},\\
d_{l m}=
&\frac{\xi_l^{(1)\prime}(\alpha)\psi_l(\alpha)-\xi_l^{(1)}(\alpha)\psi_l'(\alpha)}
{\psi_l(\bar n\alpha)\xi_l^{(1)\prime}(\alpha)-\bar n\psi_l'(\bar n\alpha)\xi_l^{(1)}(\alpha)}
B_{l m},
\end{align}
where $\bar n=n/n_\textrm{med}$ is the relative refractive index of the particle and $\xi_l^{(1)}=\psi_l-i\chi_l$, with $\chi_l$ being the Riccati-Bessel functions of the second kind. Finally, the net radiation force $\textbf{F}$ on the particle calculated using the Minkowski form of the Maxwell's stress tensor $\overleftrightarrow{T}$ can be expressed in Cartesian coordinates as
\begin{figure*}[t]
    \centering
    \includegraphics[width=0.95\textwidth]{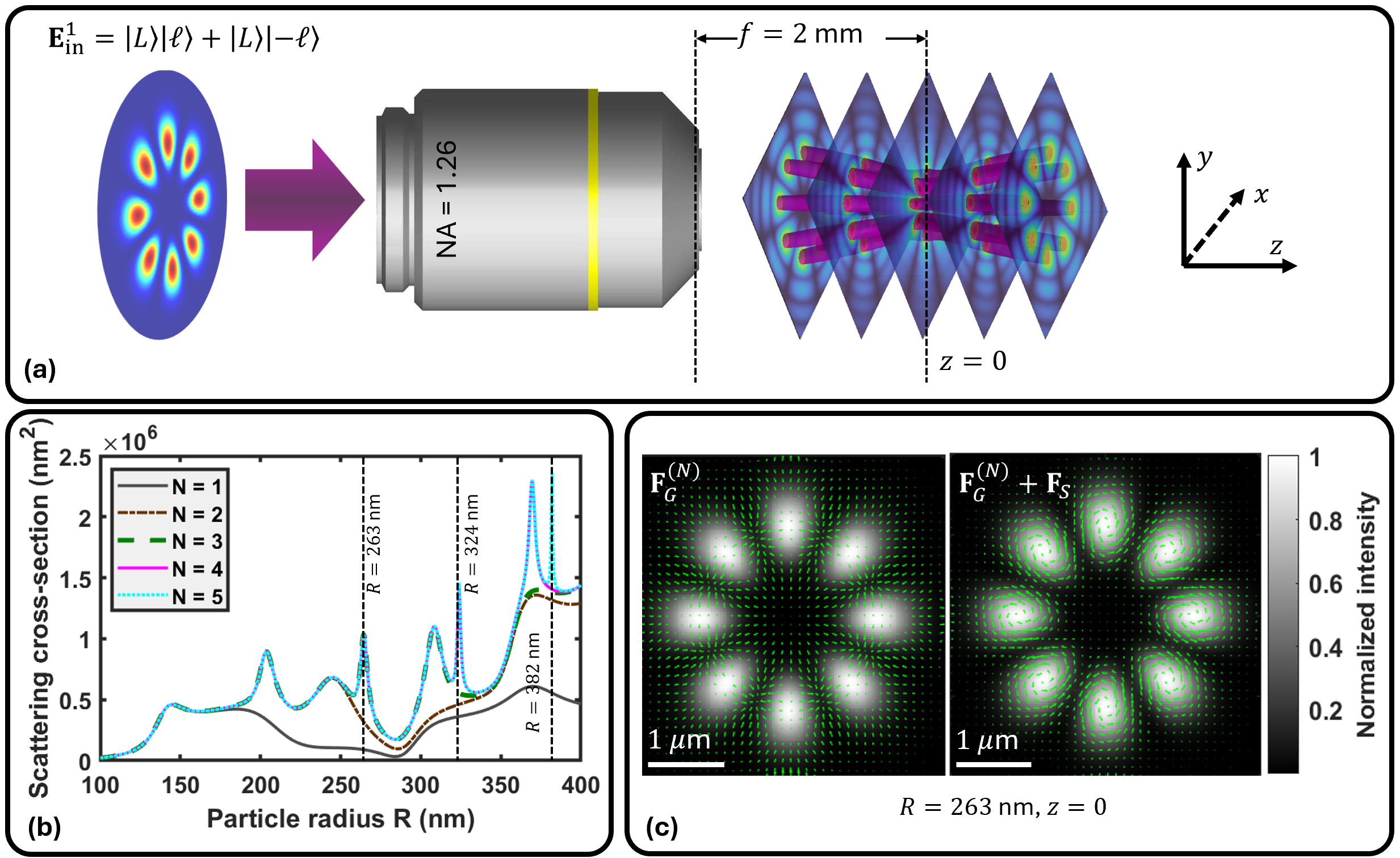} 
    \caption{Concept of multipolar optical trapping with structured cylindrical beams. (a) Schematic of the Richards-Wolf focusing of the input beam. A scalar structured beam of the form $\textbf{E}_{\textrm{in}}^1=\ket{L}\ket{\ell}+\ket{L}\ket{-\ell}$ is focused through a microscope objective of NA = 1.26, $f=2$ mm into the trapping medium (water, $n_\textrm{med}=1.33$). Multiple transverse slices of the focused beam across the focal plane $z=0$ are shown. (b) Scattering cross-section $\sigma_\textrm{sca}$ vs. particle radius $R$ line-plots for multipole truncation orders $N=1,2,...,5$. Vertical dashed lines indicate particle radii for which certain octupole ($N=3$), hexadecapole ($N=4$) and dotriacontapole ($N=5$) resonances occur. (c) Focal plane intensity distributions for $\textbf{E}_{\textrm{in}}^1$ showing gradient force vectors $\textbf{F}_G^{(3)}$ (left), and binomial force vectors $\textbf{F}_\textrm{tot}=\textbf{F}_G^{(3)}+\textbf{F}_S$ (right), for particle radius $R=263$ nm.} 
    \label{fig:Fig1}
\end{figure*}
\begin{widetext}

\begin{align}
    \frac{\langle F_x\rangle+i\langle F_y\rangle}{R^2E_0^2}
    ={}&\frac{\alpha^2}{16\pi}\,i
    \sum_{l=1}^{\infty}\sum_{m=-l}^{l}
    \Biggl(l(l+2)\,
    \sqrt{\frac{(l+m+2)(l+m+1)}
    {(2l+1)(2l+3)}}
    \notag\\[-2pt]
    &\times\Bigl(
    2\epsilon_\textrm{med} a_{l m}a^{*}_{l+1,m+1}
    +\epsilon_\textrm{med} a_{l m}A^{*}_{l+1,m+1}
    +\epsilon_\textrm{med} A_{l m}a^{*}_{l+1,m+1}
    \notag\\
    &\qquad\quad
    +2b_{l m}b^{*}_{l+1,m+1}
    +b_{l m}B^{*}_{l+1,m+1}
    +B_{l m}b^{*}_{l+1,m+1}
    \Bigr)
    \notag\\[2pt]
    &+l(l+2)\,\sqrt{\frac{(l-m+1)(l-m+2)}
    {(2l+1)(2l+3)}}
    \Bigl(
    2\epsilon_\textrm{med} a_{l+1,m-1}a^{*}_{l m}
    \notag\\
    &\qquad\quad
    +\epsilon_\textrm{med} a_{l+1,m-1}A^{*}_{l m}
    +\epsilon_\textrm{med} A_{l+1,m-1}a^{*}_{l m}
    +2b_{l+1,m-1}b^{*}_{l m}
    \notag\\
    &\qquad\quad
    +b_{l+1,m-1}B^{*}_{l m}
    +B_{l+1,m-1}b^{*}_{l m}
    \Bigr)
    \notag\\[2pt]
    &-\sqrt{(l+m+1)(l-m)}\,\sqrt{\epsilon_\textrm{med}}
    \Bigl(
    -2a_{l m}b^{*}_{l,m+1}
    +2b_{l m}a^{*}_{l,m+1}
    -a_{l m}B^{*}_{l,m+1}
    \notag\\
    &\qquad\quad
    +b_{l m}A^{*}_{l,m+1}
    +B_{l m}a^{*}_{l,m+1}
    -A_{l m}b^{*}_{l,m+1}
    \Bigr)
    \Biggr).
    \label{eq:transverse-Mie-force}
\end{align}

\begin{align}
    \frac{\langle F_z\rangle}{R^2E_0^2}
    =&-\frac{\alpha^2}{8\pi}
    \sum_{l=1}^{\infty}\sum_{m=-l}^{l}
    \operatorname{Im}\Biggl(
    l(l+2)
    \sqrt{\frac{(l-m+1)(l+m+1)}
    {(2l+3)(2l+1)}}
    \notag\\[-2pt]
    &\times\Bigl(
    2\epsilon_\textrm{med} a_{l+1,m}a^{*}_{l m}
    +\epsilon_\textrm{med} a_{l+1,m}A^{*}_{l m}
    +\epsilon_\textrm{med} A_{l+1,m}a^{*}_{l m}
    \notag\\
    &\qquad\quad
    +2b_{l+1,m}b^{*}_{l m}
    +b_{l+1,m}B^{*}_{l m}
    +B_{l+1,m}b^{*}_{l m}
    \Bigr)
    \notag\\[2pt]
    &+\sqrt{\epsilon_\textrm{med}}\,m
    \Bigl(
    2a_{l m}b^{*}_{l m}
    +a_{l m}B^{*}_{l m}
    +A_{l m}b^{*}_{l m}
    \Bigr)
    \Biggr).
    \label{eq:axial-Mie-force}
\end{align}

\end{widetext}
where $\epsilon_\textrm{med}=n_\textrm{med}^2$.
Using these equations, we can now fully determine the net radiation force on a Mie particle under arbitrary structured illumination, denoted by $\textbf{F}_\textrm{net}$ in the following sections. In particular, this prescription allows the study of tailored optical force and energy landscapes in the context of multipolar trapping. 

\section{Results and Discussion}\label{sec3}
\subsection{Landscapes by focusing cylindrical scalar beams}
To elucidate the effect of multipolar excitations on optical trapping landscape formed by spatially structured cylindrical fields, we first consider a scalar structured input beam $\textbf{E}_{\textrm{in}}^1$ at wavelength, $\lambda =1064\,$nm with azimuthal sinusoidal amplitude variation in the entrance pupil plane (c.f.\,\cite{zhang2024full}; Fig.~\ref{fig:Fig1}(a). In polar coordinates $(\rho,\phi)$, this beam is mathematically expressed as
\begin{equation}
    \textbf{E}_{\textrm{in}}^1=\frac{1}{\sqrt2}\left(\ket{L}\ket{\ell}+\ket{L}\ket{-\ell}\right), 
    \label{eq:E1}
\end{equation}
where $\ket{L}=\textbf{e}_x+\text{i}\textbf{e}_y$ denotes the left-handed circular polarization (LCP) state, and the state $\ket{\ell}$ with $\ell=4$ given by 
\begin{equation}
    \ket{\ell}=E_0\left(\frac{\rho}{w_0}\right)^{|\ell|}\exp\left(-\frac{\rho^2}{w_0^2}\right)\exp(\text{i}\ell\phi).
\end{equation}
Here, $E_0$ denotes the field amplitude corresponding to an input optical power of $200\,$mW and beam waist $w_0=0.63\,$mm. As illustrated in Fig.~\ref{fig:Fig1}(a), the input beam is focused through a microscope objective ($\text{NA} = 1.26$, $f=2\,$mm), with water ($n_\textrm{med}=1.33$) as the trapping medium (med). Note that such scalar input fields, if of circular polarization, are known to preserve the azimuthal sinusoidal amplitude variation upon tight focusing (Fig.~\ref{fig:Fig1}(a, right)), with the number of bright lobes and the beam diameter being dependent on the topological charge $\ell$. This customizable focal intensity landscape makes them an interesting candidate for bright, gray, and dark trapping with configurable arrays of trapping sites.

Owing to its high refractive index and negligible absorption at $1064\,$nm\,\cite{polyanskiy2024refractiveindex}, silicon ($n=3.56$) is chosen as the particle material for our simulations, since it is expected to enhance the multipolar Mie resonances compared to conventional silica particles\,\cite{lepeshov2023levitated,lu20263d}. To study the dependence of multipolar Mie response on the particle size, we calculate the scattering cross-section $\sigma_\textrm{sca}$ as a function of particle radius $R$ for different truncation orders ($N=1,2,...,5$) using\,\cite{bohren2008absorption}
\begin{equation}
    \sigma_\textrm{sca}=\frac{2\pi}{k^2}\sum_{l=1}^N(2l+1)(|a_l|^2+|b_l|^2).
\end{equation}
The respective results in Fig. \ref{fig:Fig1}(b) show the occurrence of sharp resonances for certain particle radii, corresponding to the excitation of higher-order ($N\geq3$) electric and magnetic multipoles. For our simulations, we discuss the results corresponding to the $R=263\,$nm octupole resonance excitation ($N=3$) and compare it to off-resonance trapping behavior for different input beams. Note that the trapping behavior will vary for the other resonance peaks at larger radii (see dashed lines in Fig.~\ref{fig:Fig1}(b)); this is discussed in brief in a subsequent section.

\begin{figure*}[t]
    \centering
    \includegraphics[width=0.95\textwidth]{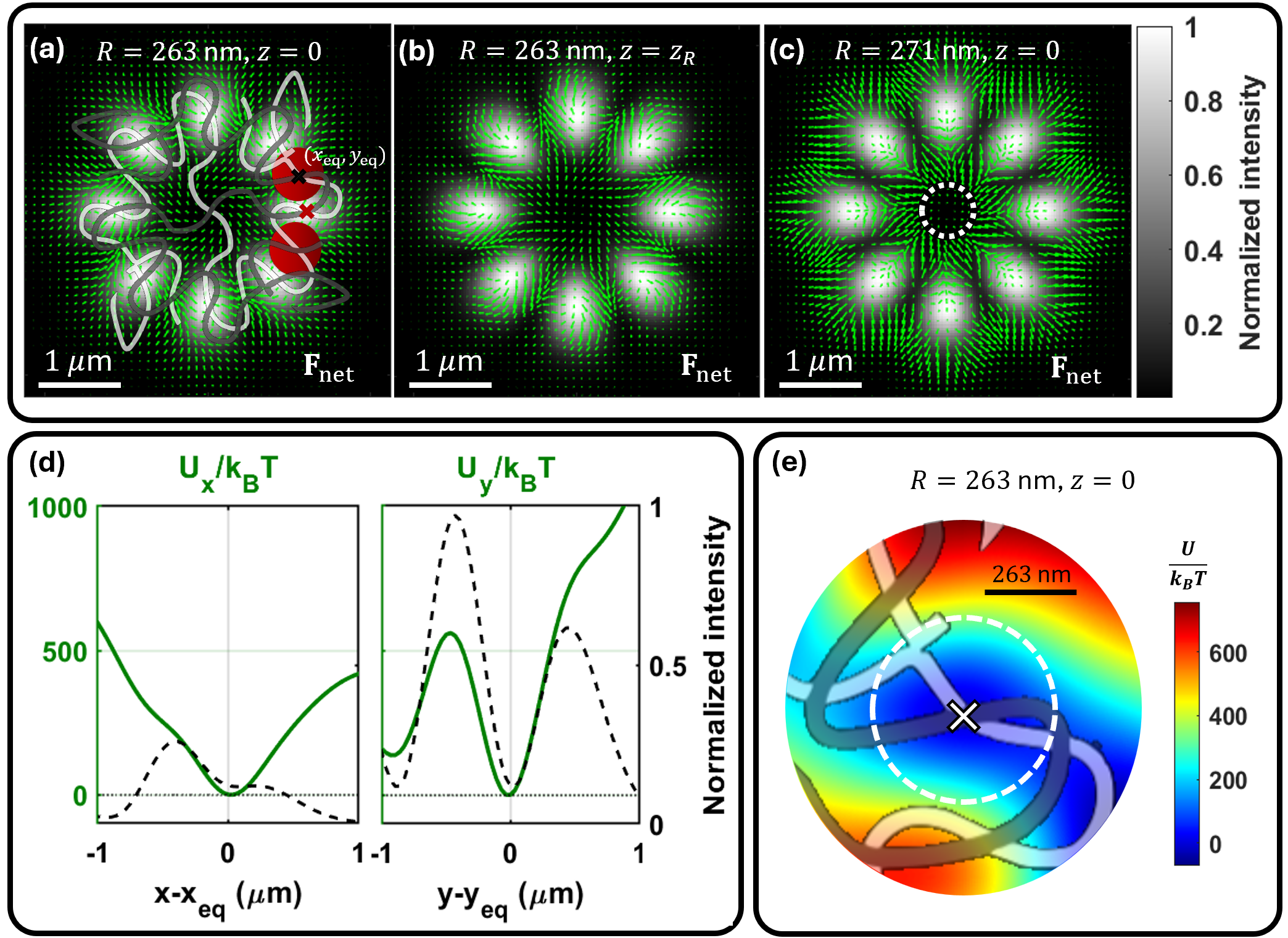} 
    \caption{Analysis of multipolar trapping for input beam $\textbf{E}_{\textrm{in}}^1$ (transverse plane). (a) Focal plane intensity distribution with net radiation force vectors $\textbf{F}_\textrm{net}$ calculated using GLMT for $R=263$ nm. Light and dark gray overlaid plots show the zerolines of the $x,\,y$-force components. Black cross marks the zerolines intersection corresponding to stable equilibrium $(x_\textrm{eq},y_\textrm{eq})$; red cross marks unstable equilibrium. The two red spheres are representative of the gray-trapped silicon particles, indicating their size in relation to the trapping landscape. (b) Intensity distribution with $\textbf{F}_\textrm{net}$ calculated at $z=z_R\approx284\,$nm for $R=263\,$nm. (c) Focal plane intensity distribution with $\textbf{F}_\textrm{net}$ calculated for $R=271\,$nm illustrating (off-resonance) dark trapping at the center. The dotted circle marks the particle boundary. (d) Potential energies $U_x/(k_\textrm{B}T)$ and $U_y/(k_\textrm{B}T)$ calculated along $x$ and $y$-axes respectively around $(x_\textrm{eq},y_\textrm{eq})$ (solid green) plotted with normalized intensity (dashed black). (e) Radial potential energy centered at stable equilibrium $(x_\textrm{eq},y_\textrm{eq})$, showing zerolines of the force intersecting at the equilibrium point. Dashed white circle marks the particle boundary.}
    \label{fig:Fig2}
\end{figure*}
First let us illustrate the significance of choosing the appropriate theoretical description in such complex trapping scenarios. Figure~\ref{fig:Fig1}(c, left) shows only the gradient force (green vectors on focal intensity) acting on a particle of radius $263\,$nm in the focal plane of the input beam $\textbf{E}_{\textrm{in}}^1$, calculated using Eq. \ref{eq: gradient-force}. The convergence of the gradient force vectors around the regions between the bright petals indicates the possibility of transverse gray trapping (c.f. Ref.~\cite{zhang2024full}). In contrast, the total force (in Fig.~\ref{fig:Fig1}(c, right)), calculated using Eq.~(\ref{eq: grad_plus_scatt_force}), hence following the binomial representation, shows the force vectors spiraling into the bright regions; this observation contradicts the prospects of gray trapping. To resolve the contradiction, the net radiation force $\textbf{F}_\textrm{net}$ is calculated at the focal plane ($z=0$) using Eq.~(\ref{eq:transverse-Mie-force}), i.e. GLMT. Note that, in our formulation, $\textbf{F}_\textrm{net}$ is distinct from $\textbf{F}_\textrm{tot}$ defined earlier (Eq.\,\ref{eq: grad_plus_scatt_force}). The results are presented in Fig.~\ref{fig:Fig2}(a). The force vectors can be found pointing towards the gray intensity regions in agreement with the similar gray trapping results in Ref.~\cite{zhang2024full}. 

We identify regions of gray trapping by finding the intersections of the zerolines of $(\textbf{F}_\textrm{net})_x$ and $(\textbf{F}_\textrm{net})_y$ (see light/dark gray lines in Fig.~\ref{fig:Fig2}(a)). The force Jacobian (stiffness matrix\,\cite{kiselev2016optical}), yielding two negative eigenvalues, implies a stable equilibrium at zeroline intersections between high-intensity lobes (exemplarily marked with a black cross, $(x_\textrm{eq},y_\textrm{eq})$ in Fig. \ref{fig:Fig2}(a)). In contrast, at least one positive eigenvalue of the Jacobian implies an unstable equilibrium or a saddle point, as exemplarily marked with the red cross (zeroline intersections at intensity maxima). To confirm the evidence for stable gray trapping between the intensity lobes for $R=263\,$nm, we assess the potential energy distribution using the corresponding work integrals $\int(\textbf{F}_\textrm{net})_xdx$ and $\int(\textbf{F}_\textrm{net})_ydy$ calculated around $(x_\textrm{eq},y_\textrm{eq})$ (Fig.~\ref{fig:Fig2}(d)), and the radial work integral $\int(\textbf{F}_\textrm{net})_rdr$ (Fig. \ref{fig:Fig2}(e)). These calculations quantify a potential energy depth exceeding 400 $k_\textrm{B}T$ at $(x_\textrm{eq},y_\textrm{eq})$, confirming stable transverse confinement. Further, the transverse optical forces ($\sim10^{-11}\,$N) are significantly stronger than the corresponding gravitational force ($\sim10^{-15}\,$N) on the particle.

For comparison, we repeat our calculations at the focal plane for the (off-resonance) radius of $R=271\,$nm, as illustrated in Fig.~\ref{fig:Fig2}(c). While we can no longer find converging force vectors between intensity petals for this particle radius, we find them at the center of the beam. Following the explanation for a similar trapping field by Zhang et al.~\cite{zhang2024full}, dark trapping could be expected in this central area due to the interference between the multipolar excitations reversing the direction of gradient force away from the intensity maxima. Note that although the distribution of the $\textbf{F}_\textrm{net}$ vectors immediately suggests dark trapping stability, the magnitude of the transverse net radiation force close to the center ($\sim10^{-16}\,$N) is not strong enough to overcome the downward ($-y$ direction) gravitational pull ($\sim10^{-15}\,$N) on the particle in the more common horizontal configuration of the trapping setup (cf. Fig.~\ref{fig:Fig1}(a)). Hence, we do not expect stable, off-resonance dark trapping in this scenario.

Usually, stable trapping is achieved close to the focal plane, i.e., $z=0$; this position and the stability of the trap are, however, dependent on the longitudinal evolution of the trapping field. To gain first insights into this longitudinal evolution and respective effects on the (resonant) trapping efficiency for $R=263\,$nm, we calculate $\textbf{F}_\textrm{net}$ at the plane $z=z_R$, as shown in Fig.\,\ref{fig:Fig2}(b). Here, $z_R\approx284\,$nm corresponds to the Rayleigh range of a Gaussian beam of the same beam waist as $\textbf{E}_{\textrm{in}}^1$ at the focal plane. While the force vector distribution has slightly changed in comparison to $z=0$, the potential energy calculations still yield potential depth around 100 $k_\textrm{B}T$, suggesting decreased yet sufficient stability for achieving transverse gray trapping away from $z=0$ plane. 

\begin{figure}[h!]
    \centering
    \includegraphics[width=0.9\columnwidth]{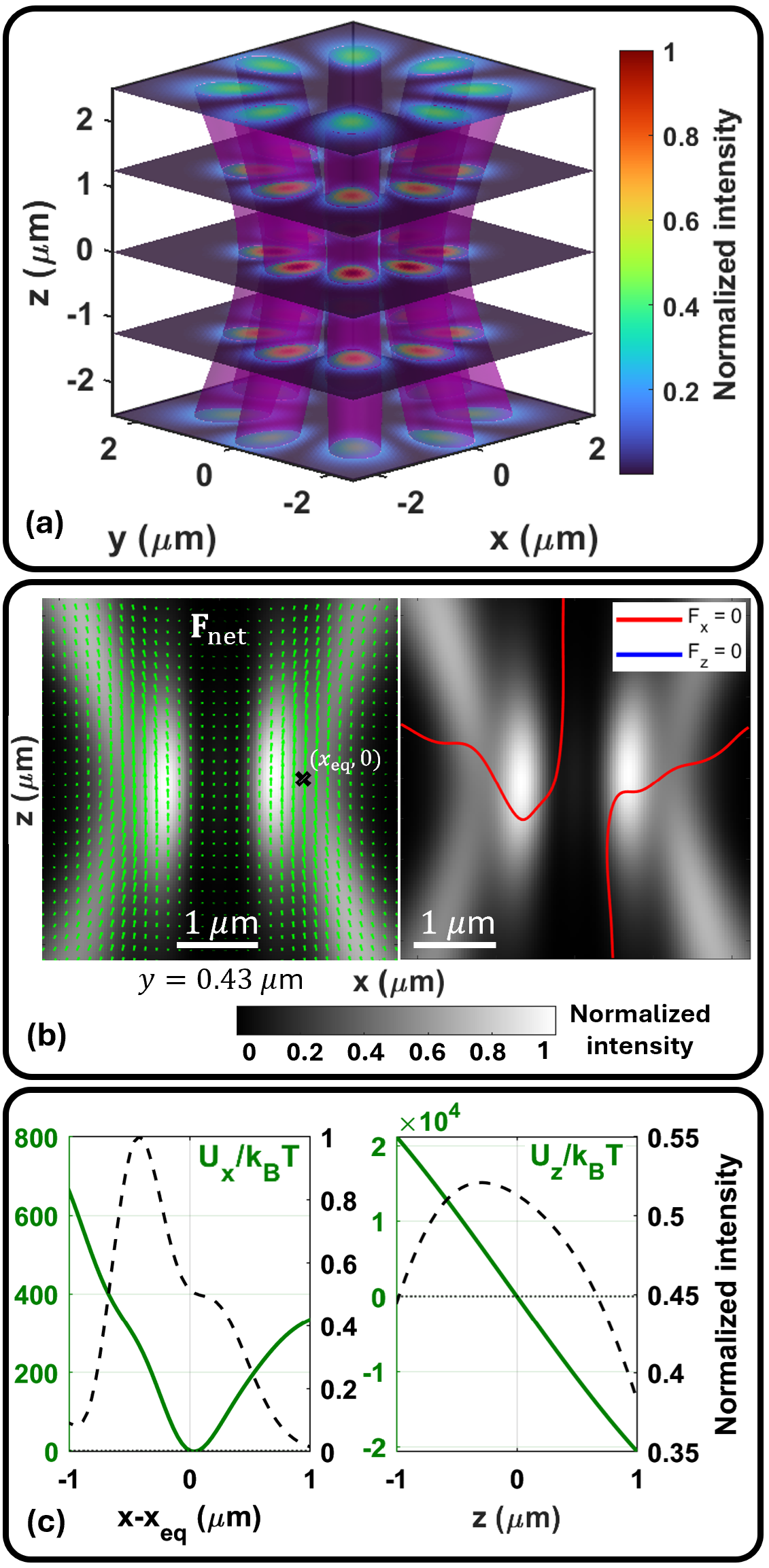} 
    \caption{Analysis of multipolar axial trapping for input beam $\textbf{E}_{\textrm{in}}^1$. (a) 3D intensity distribution in the focal region showing multiple transverse planes across $z=0$ with the magenta tubes showing the $20\%$ intensity isosurfaces. (b) Axial ($xz$-plane) intensity distribution with $\textbf{F}_\textrm{net}$ vectors calculated at $y=0.43\,\mu$m (left), and zerolines of the axial $(xz)$ force (right). The point marked with a black cross $(x_\textrm{eq},0)$ corresponds to the location where potential energy is calculated. (c) Work integrals for the $x$ and $z$ components (left and right, respectively) of the net radiation force calculated around the black cross $(x_\textrm{eq},0)$. A stable equilibrium can be found in $x$ but not in $z$ direction.}
    \label{fig:Fig3}
\end{figure}

We further investigate the longitudinal field evolution and, hence, the axial trapping scenario by extending our calculations to three dimensions (3D). We illustrate the 3D intensity distribution for $\textbf{E}_{\textrm{in}}^1$ in Fig.\,\ref{fig:Fig3}(a). The $z$-evolution of the intensity isosurfaces (in magenta) around the focal plane suggests relatively weak axial focusing. This is reflected in the net radiation force distribution and the zerolines of the force in the $xz$-plane (calculated at transverse equilibrium $y=0.43\,\mu$m using Eq. \ref{eq:axial-Mie-force}), illustrated in Fig. \ref{fig:Fig3}(b). The absence of $F_z$ zerolines confirms the absence of a stable axial equilibrium. As shown in Fig.~\ref{fig:Fig3}(c, right), this claim is further substantiated by the potential energy distribution in $z$ direction being monotonic around the transverse equilibrium position $(x_\textrm{eq},0)$. 

Hence, while the transverse forces alone indicate stable resonant gray trapping, our 3D analysis reveals challenging axial trapping behavior; for off-resonant dark trapping, the dominating gravitational force prohibits the desired stability in a horizontal trapping scenario. To tackle these challenges, we advance from scalar input light fields to trapping beams of spatially varying input polarization, namely, higher-order CVBs. Such beams embed a spatially varying ratio of azimuthal vs. radial polarization components, which allows for sophisticated 3D-polarized focal fields and stronger focal field confinement\,\cite{youngworth2000focusing, Otte2017FocalLandscapes}. These properties could enhance net radiation forces, that could help overcome the gravitational force. We investigate the trapping characteristics of higher-order CVBs in the following section.

\subsection{Landscapes by focusing cylindrical vector beams}
\begin{figure}[htbp]
    \centering
    \includegraphics[width=\columnwidth]{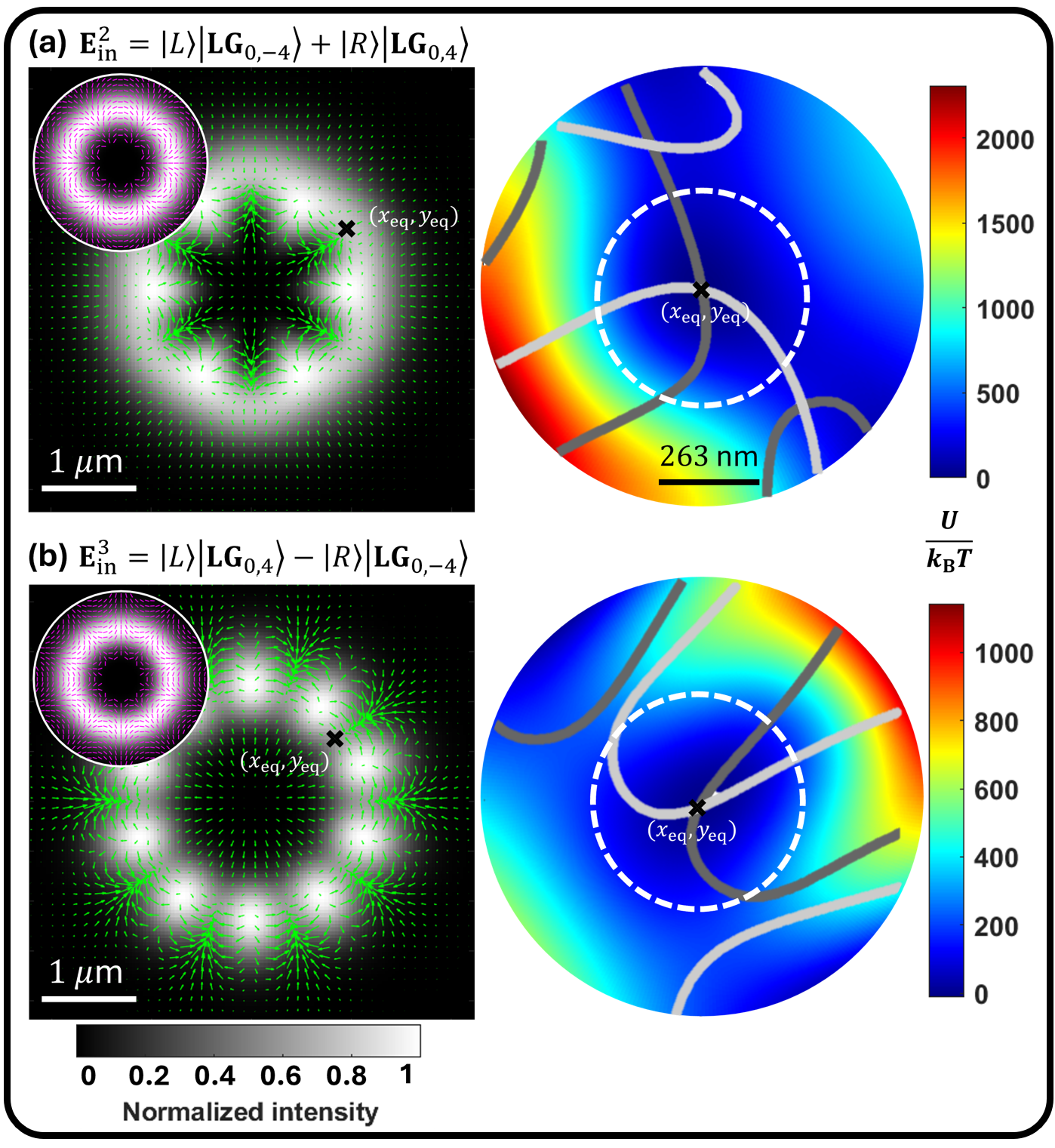} 
    \caption{Multipolar trapping with focused cylindrical vector beams ($p=0$). (a) (Left) Focal plane intensity distribution for input beam $\textbf{E}_{\textrm{in}}^2$ ($p=0,\ell=4$) with $\textbf{F}_\textrm{net}$ vectors (green). Black cross indicates the point $(x_\textrm{eq},y_\textrm{eq})$ where stable equilibrium occurs. Inset: Intensity at the entrance pupil showing six-petals-flower polarization structure (magenta). (Right) Radial potential energy $U/(k_\textrm{B}T)$ centered at $(x_\textrm{eq},y_\textrm{eq})$ coincident with the intersection of force zerolines (light and dark gray lines). Dashed white circle marks the particle boundary. (b) (Left) Focal plane intensity distribution for input beam $\textbf{E}_{\textrm{in}}^3$ ($p=0,\ell=4$) with $\textbf{F}_\textrm{net}$ vectors (green). Inset: Intensity at the entrance pupil showing ten-sectors-web polarization structure (magenta). (Right) Corresponding radial potential energy with zerolines and particle boundary.}
    \label{fig:Fig4}
\end{figure}
We explore gray and dark trapping prospects for two representative input CVBs $\textbf{E}_{\textrm{in}}^2$ and $\textbf{E}_{\textrm{in}}^3$, which can be represented in the circular polarization basis $\{\ket{L},\ket{R}\}$ as
\begin{align}
    \textbf{E}_{\textrm{in}}^2=\frac{1}{\sqrt2}\left(\ket{L}\ket{\textrm{LG}_{p,-\ell}}+\ket{R}\ket{\textrm{LG}_{p,\ell}}\right), \label{eq:E2}\\
    \textbf{E}_{\textrm{in}}^3=\frac{1}{\sqrt2}\left(\ket{L}\ket{\textrm{LG}_{p,\ell}}-\ket{R}\ket{\textrm{LG}_{p,-\ell}}\right),\label{eq:E3}
\end{align}
with $\ket{\textrm{LG}_{p,\ell}}$ representing a Laguerre-Gaussian mode of radial index $p\in \mathbb{N}_0$ and topological charge $\ell$. In polar coordinates $(\rho, \phi, z)$, the complex amplitude of $\ket{\textrm{LG}_{p,\ell}}$ is given by\,\cite{siegman1986lasers, Boyd1961}

\begin{align} 
    & \text{LG}_{p,\ell}(\rho,\,\phi,\,z) = A_{p,\ell}(\rho,z) \cdot \text{e}^{\text{i} \frac{k \rho^2}{2R'(z)}}\cdot \text{e}^{\text{i}\phi_{p,l}^{G}(z)}\cdot \text{e}^{\text{i}\ell \phi}, \label{eq:LGbeam}\\
    &A_{p,\ell}(\rho,z) = E_0\sqrt{\frac{2p!}{\pi(|\ell|+p)!}}\cdot \frac{1}{w(z)}\cdot \text{e}^{-\frac{\rho R'(z){^2}}{w(z)^2}} \\ \notag
    & \qquad \qquad \qquad \cdot \left(\frac{\rho\sqrt{2}}{w(z)}\right)^{|\ell|}\cdot \text{L}_p^{|\ell|}\left(\frac{2\rho^2}{w(z)^2}\right),\\ \notag
    &\phi_{p,\ell}^G(z) = (2p+|\ell|+1) \, \phi_{0,0}^G(z).
\end{align}
In these equations, $R'(z)$ is the wave front curvature, $w(z)$ the beam radius ($w_0=w(0)$: beam waist),  $\text{L}_p^\ell(\cdot)$ represents the eponymous Laguerre polynomial, $E_0$ the field amplitude, and $\phi_{p,\ell}^G$ the Gouy phase shift ($\phi_{0,0}^G$: Gouy phase of fundamental Gaussian beam).

Note that, while for $\textbf{E}_{\textrm{in}}^1$ (Eq.~(\ref{eq:E1})), we combined two equally polarized beams carrying opposite phase vortices $\exp(\pm\text{i}\ell \phi)$, we now polarize the combined vortex beams orthogonally circular (Eq. (\ref{eq:E2}, \ref{eq:E3})). Hence, instead of an azimuthally varying intensity for the input field, this combination yields donut shaped intensity structures with an azimuthally varying ratio of radial and azimuthal polarization components -- so-called flower or web polarization structures\,\cite{otte2016higher}.

We start with the simpler case of $p=0$ and $\ell=4$ (same topological charge as for our $\textbf{E}_{\textrm{in}}^1$ example) corresponding to a single donut-shaped intensity distribution at the entrance pupil of the microscopy objective; we consider a particle of radius $R=263\,$nm.  
Figure~\ref{fig:Fig4}(a) illustrates the focal intensity distribution along with the net radiation force for an input CVB with six-petal-flower polarization structure ($\textbf{E}_{\textrm{in}}^2$, see inset with magenta polarization states on intensity). The force vectors localized around the vertices of the star-shaped intensity structure suggest six potential gray trapping locations, with one such equilibrium position $(x_\textrm{eq},y_\textrm{eq})$ marked with a black cross. Note that the number of vertices and therefore the potential trapping locations is directly related to the topological charge of the input field\,\cite{Otte2017FocalLandscapes}.
\begin{figure*}[t]
    \centering
    \includegraphics[width=0.85\textwidth]{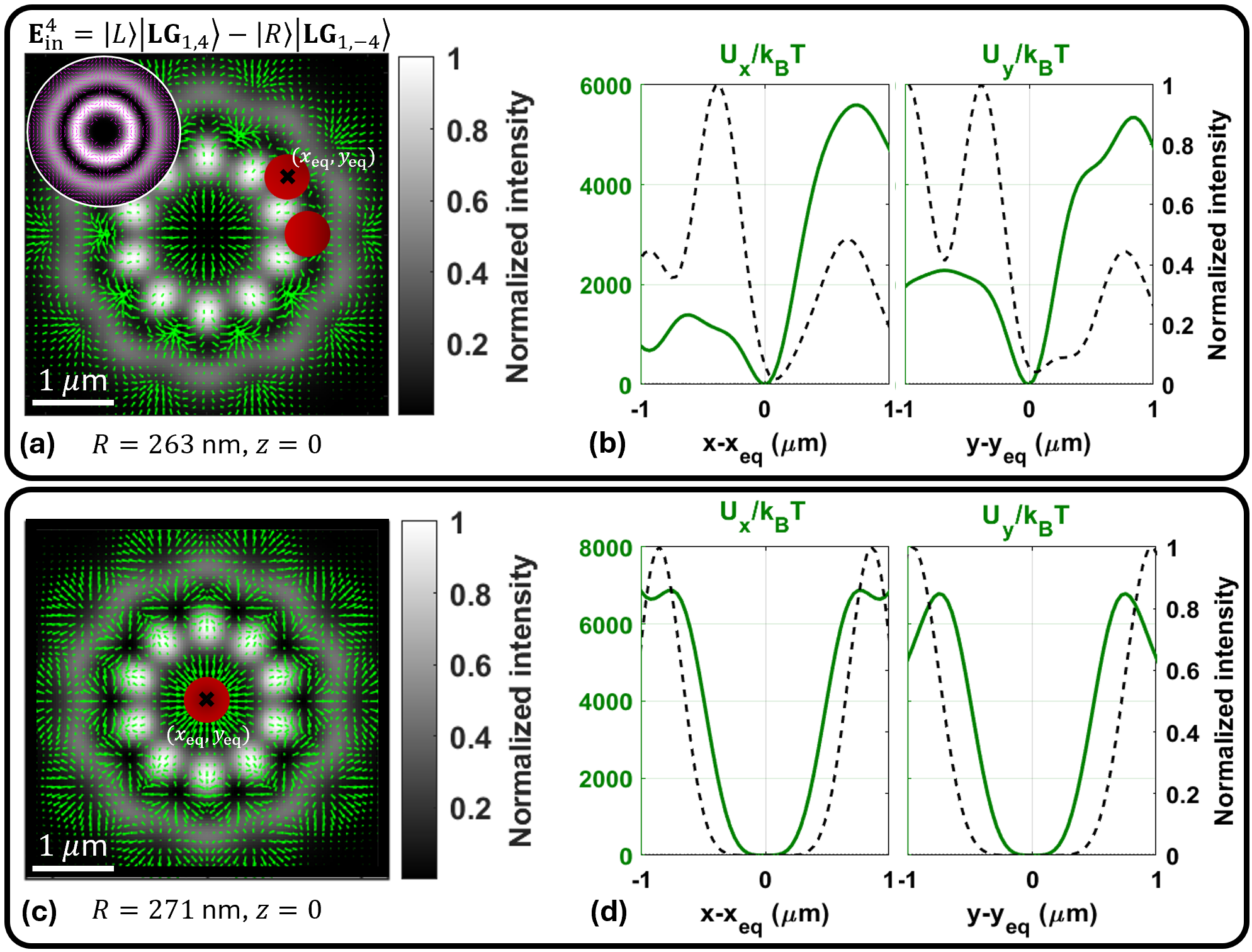} 
    \caption{Multipolar trapping with a focused higher-radial-order CVB, namely, input beam $\textbf{E}_{\textrm{in}}^3$ ($n=1,\ell=4$). (a) Focal plane intensity distribution showing transverse $\textbf{F}_\textrm{net}$ vectors (green) for $R=263$ nm. Red spheres represent two possible stable gray trapping locations with an exemplary position $(x_\textrm{eq},y_\textrm{eq})$ marked with a black cross. Inset: Input beam's intensity and (magenta) polarization distributions showing two bright rings and ten-fold web-polarization patterns. (b) Corresponding work integrals along $x$ and $y$ around $(x_\textrm{eq},y_\textrm{eq})$. (c) Focal plane intensity distribution showing transverse $\textbf{F}_\textrm{net}$ vectors for $R=271$ nm (off-resonance). Marked black cross at the center indicates stable dark trapping. (e) Corresponding potential energy at the center showing potential depth $\Delta U>6000\,k_\textrm{B}T$.}
    \label{fig:Fig5}
\end{figure*}
The potential energy plot (right) shows the radial work integral corresponding to the equilibrium position. Similar calculations are performed for the second input beam $\textbf{E}_{\textrm{in}}^3$ with ten-sectors-web polarization structure, with the results illustrated in Fig.~\ref{fig:Fig4}(b). Note that this structure upon focusing allows for ten gray trapping regions located between the bright lobes; again, the number of bright lobes and therefore trapping locations is dependent on the topological charge $\ell$ of the input\,\cite{Otte2017FocalLandscapes}. The corresponding radial potential energy plot (right) quantifies the energy depth the particle would experience around the marked equilibrium position $(x_\textrm{eq},y_\textrm{eq})$. 

For both the beams under consideration, we achieve potential energy depths $\Delta U\sim1000\,k_\textrm{B}T$, confirming the stability of transverse optical trapping in these gray-intensity regions. However, with the same input optical power and underlying beam waist as the scalar cylindrical input beam $\textbf{E}_\textrm{in}^1$, we encounter similar challenges for dark trapping at the center, where the radiation forces at the center of the focal plane fail to overcome the gravitational pull. Since these CVBs offer flexibility in designing configurable arrays of trapping sites with stable potential wells, a possible remedy is to boost the input optical power if practical in the experimental system at hand. 

For potentially overcoming the aforementioned challenges, we leverage the so far unexplored radial parameter $p$ of the input CVB. For $p\neq 0$, helical LG beams and, hence, CVBs formed by them show a multi-ring intensity structure with $p+1$ rings. The respective focal field inherits this multi-ring structure, potentially enabling increased transverse field gradients with stronger radial and azimuthal confinement of the intensity distribution, which could unlock stronger net radiation forces in the focal plane. This motivates the shift from $p=0$ to higher radial-order CVBs. To explore this hypothesis, let us investigate the example of a tightly focused $\textbf{E}_{\textrm{in}}^3$ beam with $p=1,\ell=4$ (henceforth denotes as $\textbf{E}_{\textrm{in}}^4$) as the trapping field. We present the transverse radiation force landscape (green force vectors on intensity) acting on a particle of $R=263\,$nm in Fig.~\ref{fig:Fig5}(a); we show the two-ring intensity distribution at the entrance pupil overlaid with the tens-sector-web polarization pattern (magenta) as an inset. The regions of gray trapping can be found to occur between the bright lobes (of the inner ring) and bound between the two rings of the intensity distribution, allowing for up to ten potential locations for gray trapping. The potential energy distributions along $x$ and $y$ for one such transverse plane location $(x_\textrm{eq},y_\textrm{eq})$ (black cross) is illustrated in Fig.~\ref{fig:Fig5}(b), showing energy depths exceeding 1000 $k_\textrm{B}T$. Note that (ten) additional stable gray trapping sites are found to occur outside the outer ring of the focal intensity, albeit of lower potential energy depths ($\sim100\,k_\textrm{B}T$). 

Finally, to explore the full capabilities of higher-order CVBs for multipolar optical trapping, we return to the off-resonant trapping regime. We perform the same calculations as above for $R=271\,$nm with the results illustrated in Fig.~\ref{fig:Fig5}(c),(d). The distribution of transverse $\textbf{F}_\textrm{net}$ and potential energy depth exceeding $6000\,k_\textrm{B}T$ indicate stable dark trapping at the center of the focal plane. The corresponding net radiation force ($\sim10^{-13}$ N) is now two orders of magnitude stronger than the gravitational force. Crucially, this differential trapping behavior based on particle size -- precisely, gray trapping for $R=263\,$nm between the bright lobes of the inner intensity ring, versus dark central trapping for $R=271\,$nm -- could find applications in particle sorting\,\cite{bobkova2021optical,kong2025customized,otte2020optical} with nanometer-level precision. In addition, such trapping field configurations could facilitate creating denser particle arrays for trapping-assisted nanopatterning. 

We emphasize two key improvements with the higher-radial-order CVB $\textbf{E}_{\textrm{in}}^4$ over the scalar structured beam $\textbf{E}_{\textrm{in}}^1$ considered earlier. First, with the net radiation force at the center of the focal plane being stronger than the gravitational force, this CVB can enable stable on-axis dark trapping trapping in the horizontal trapping configuration. Secondly, not only does this beam provide the desired improved stability, namely, more than twice the trapping stability (measured by potential energy depth) as the $\textbf{E}_{\textrm{in}}^1$ beam, but also enables the neighboring trapping sites to be pushed closer to one another, with inter-particle (center-to-center) distances around $700\,$nm. Combined with the strong field confinement offered by the focused CVB, this could provide an ideal platform for studying optical binding forces under enhanced scattering conditions at the mesoscopic scale. 

A more holistic analysis pertaining to the higher-radial-order CVB requires further investigation of its 3D trapping capabilities. To this end, we studied the axial trapping behavior for input beam $\textbf{E}_{\textrm{in}}^4$. The analysis still bears similar results as the scalar structured beam, implying unviability of a single-beam trapping configuration. This can however be mitigated by different approaches, such as using electrically charged silicon spheres and trapping them against an oppositely charged glass slide\,\cite{li2025optical}. Another workaround could be using anti-reflection-coated (ARC) particles for trapping to reduce the axial scattering forces\,\cite{wang2018gradient}, but this might compromise the geometry of the multipolar excitations in the particle and consequently the gray and dark trapping behavior. We also investigated the focal plane GLMT force distributions corresponding to the other higher-order multipolar resonances -- $R=324,\,382$ nm, for the three input beams $\textbf{E}_\textrm{in}^{1}$, $\textbf{E}_\textrm{in}^{2}$, and $\textbf{E}_\textrm{in}^{3}$. The force distributions show distinct patterns at different particle radii, however, their capabilities in the context of stable optical trapping are inconclusive within the scope of this study and needs further examination.

\section{Conclusion}\label{sec4}
We have shown theoretically that the interplay between structured illumination and resonant multipolar scattering enables control over the location and geometry of transverse optical equilibria beyond intensity maxima. For silicon spheres near an octupolar resonance, generalized Lorenz–Mie calculations predict gray trapping that the phenomenological gradient-plus-scattering prescription fails to reproduce. We compared (cylindrical) scalar and vector structured beams and found that the flower- and web-polarization-structured vector beams showed better adaptability in organizing gray trapping sites into distinct spatial arrangements. We predicted that a higher-radial-order web-polarization beam supports closely spaced off-axis gray trapping equilibria and central dark confinement for distinct particle radii. For these trapping sites, along the evaluated transverse paths, the calculated optical force integrals exceed $10^3$ $k_\textrm{B}T$ for gray confinement and $6\times10^3$ $k_\textrm{B}T$ for dark confinement. These findings establish beam polarization, mode structure, and particle size as complementary parameters for tailoring multipolar force landscapes. The predicted radius-dependent equilibria and approximately $700\,$nm separation between neighboring gray trapping sites offer prospects for size-selective particle organization, trapping assisted nanopatterning, and investigations of optical binding between resonant particles. With stable transverse confinement and configurable trapping landscapes, our approach potentially also unravels applications in non-invasive manipulation of sub-micron-scale biomolecules. Our work takes a step forward in understanding the interplay between the non-paraxial 3D polarization of the trapping beam and the resonant excitation of multipoles in the trapped particle. Extending our present single-particle description to include multiple scattering and collective dynamics could be a promising avenue for future work.   

\section{Conflict of Interest}
The authors declare no conflicts to disclose.


\section{Acknowledgment}\label{sec7}
N.M. acknowledges support from the National Science Foundation (NSF, Award Number 2513310).

\bibliography{ref}
\bibliographystyle{ieeetr}

\end{document}